\documentclass[twocolumn,floatfix,superscriptaddress,a4paper,showpacs,showkeys,nofootinbib,reprint,prc]{revtex4-1}

\usepackage{epsfig}
\usepackage{latexsym}
\usepackage{xspace}
\usepackage[colorlinks=true,linktocpage=true,linkcolor=blue,citecolor=blue,allcolors=blue]{hyperref}
\usepackage[utf8]{inputenc}
\usepackage{indentfirst}
\usepackage{enumerate}
\usepackage{color}
\usepackage{tabularx}

\usepackage{amsmath}
\usepackage{amssymb}
\usepackage[english]{babel}
\usepackage{url}
\newcommand{\eq}[1]{\begin{align} #1 \end{align}}

\begin{document}


\title{
Multiplicity dependence of light nuclei production at LHC energies\\
in the canonical statistical model
}

\author{Volodymyr Vovchenko}
\affiliation{
Institut f\"ur Theoretische Physik,
Goethe Universit\"at Frankfurt, Max-von-Laue-Str. 1, D-60438 Frankfurt am Main, Germany}
\affiliation{Frankfurt Institute for Advanced Studies, Giersch Science Center, Goethe Universit\"at Frankfurt, Ruth-Moufang-Str. 1, D-60438 Frankfurt am Main, Germany}

\author{Benjamin D\"onigus}
\affiliation{
Institut f\"ur Kernhysik,
Goethe Universit\"at Frankfurt, Max-von-Laue-Str. 1, D-60438 Frankfurt am Main, Germany}

\author{Horst Stoecker}
\affiliation{
Institut f\"ur Theoretische Physik,
Goethe Universit\"at Frankfurt, Max-von-Laue-Str. 1, D-60438 Frankfurt am Main, Germany}
\affiliation{Frankfurt Institute for Advanced Studies, Giersch Science Center, Goethe Universit\"at Frankfurt, Ruth-Moufang-Str. 1, D-60438 Frankfurt am Main, Germany}
\affiliation{
GSI Helmholtzzentrum f\"ur Schwerionenforschung GmbH, Planckstr. 1, D-64291 Darmstadt, Germany}

\begin{abstract}
The statistical model with exact conservation of baryon number, electric charge, and strangeness -- the Canonical Statistical Model (CSM) --
is used to analyze the dependence of yields of light nuclei at midrapidity on charged pion multiplicity at the LHC.
The CSM calculations are performed assuming baryon-symmetric matter, using the recently developed \texttt{Thermal-FIST} package.
The light nuclei-to-proton yield ratios show a monotonic increase with charged pion multiplicity, with a saturation at the corresponding grand-canonical values in the high-multiplicity limit,
in good qualitative agreement with the experimental data measured by the ALICE collaboration in pp and Pb-Pb collisions at different centralities and energies.
Comparison with experimental data at low multiplicities shows that exact conservation of charges across more than one unit of rapidity and/or a chemical freeze-out temperature which decreases with the charged pion multiplicity improves agreement with the data.
\end{abstract}

\pacs{24.10.Pa, 25.75.Gz}

\keywords{statistical model, light nuclei production, canonical suppression}

\maketitle



Relative hadron yields measured in heavy-ion collisions at various energies are known to be described surprisingly well by the thermal-statistical model~\cite{Cleymans:1992zc,BraunMunzinger:1996mq,Becattini:2000jw}, which  in the simplest case represents a non-interacting gas of known hadrons and resonances in the grand canonical ensemble~(see, e.g., Ref.~\cite{Braun-Munzinger:2015hba} for an overview).
This concept has also been long applied to light nuclei~\cite{Mekjian:1977ei,Mekjian:1978us,Mekjian:1978zz,Gosset:1988na,Siemens:1979dz,Stoecker:1984py,Hahn:1986mb,Csernai:1986qf,Jacak:1987zz,Hahn:1988zz}, and, even more surprisingly, a very good description of the various light (hyper)nuclei yields measured in heavy-ion collisions is obtained~\cite{BraunMunzinger:1994iq,Andronic:2010qu,Steinheimer:2012tb,Adam:2015vda,Adam:2015yta,Anticic:2016ckv}. 
This is so despite the small binding energies of these nuclei~(of the order of few MeV or less) relative to the typical chemical freeze-out temperatures $T_{\rm ch} \sim 150$~MeV.

In the standard chemical equilibrium grand-canonical description all conserved charges are only conserved ``on average'', but fluctuate from one microscopic state to another.
This grand-canonical treatment of particle yields is appropriate when the reaction volume is sufficiently large.
However, when the reaction volume is small, i.e. when the number of particles with particular conserved charge(s) is of the order of unity or smaller, then the canonical treatment of the corresponding conserved charge(s) is necessary~\cite{Rafelski:1980gk,Hagedorn:1984uy,Cleymans:1990mn}.
In the canonical ensemble the conserved charges are conserved exactly, from one microscopic state to another, which results in the so-called canonical suppression of the yields of particles carrying conserved charges, relative to their grand-canonical values. The effect is stronger for  multi-charged particles, such as multi-strange hyperons or light nuclei.
The canonical ensemble formulation of the statistical model -- the CSM -- has been successfully used to describe hadron abundances measured in small systems, including those created in such 'elementary' collisions as $e^+ e^-$~\cite{Becattini:1995if,Andronic:2008ev,Becattini:2008tx}, pp or p($\bar{\text{p}}$)~\cite{Becattini:1997rv,Becattini:2005xt,Becattini:2010sk}.

Canonical suppression effects have previously been considered at LHC energies, but either for strangeness only~\cite{Kraus:2008fh,Adam:2015vsf,Vislavicius:2016rwi} or without light nuclei~\cite{Sharma:2018uma}.
A qualitative description of the multiplicity dependence of ratios of yields of various strange hadrons to pions was obtained in the strangeness-canonical ensemble picture~\cite{Adam:2015vsf,Vislavicius:2016rwi}.
In the present work we consider the full canonical treatment of baryon number, electric charge, and strangeness, and include both hadrons and light nuclei.
All three conserved charges can be expected to influence the yields of light nuclei, especially the baryon number, given that light nuclei carry multiple baryon charges.
To the best of our knowledge, no such study has been performed before for the yields of light nuclei.

It should be noted that the mechanism for light nucleus formation is currently debated between thermal and coalescence~\cite{Butler:1961pr,Butler:1963pp,Csernai:1986qf} approaches.
They are based on different assumptions but both often give very similar predictions.
This issue is not discussed in the present work, but  the effects of canonical suppression are rather explored for the  thermal production mechanism scenario.
The validity of the point-particle approximation, for light nuclei in the thermal model approach, can similarly be questioned.
The excluded-volume corrections have significant effects on the thermal calculations of particle yields at LHC energies for both, hadrons and light nuclei~\cite{Vovchenko:2015cbk,Vovchenko:2016mwg}. 
This issue will be considered in a forthcoming paper.

We restrict our considerations to the ideal hadron resonance gas (HRG) in the Boltzmann approximation and in full chemical equilibrium.
In the canonical ensemble the three abelian charges considered -- the baryon number $B$, the electric charge $Q$, and the strangeness $S$ -- are fixed exactly to particular values which are conserved exactly across the so-called correlation volume $V_c$.
The partition function of the HRG model, in the canonical ensemble at a given temperature $T$ and correlation volume $V_c$, reads~\cite{Becattini:1995if,Becattini:1997rv}
\eq{\label{eq:Z}
\mathcal{Z}(B,Q,S) & =
\int \limits_{-\pi}^{\pi}
  \frac{d \phi_B}{2\pi}
 \int \limits_{-\pi}^{\pi}
  \frac{d \phi_Q}{2\pi}
  \int \limits_{-\pi}^{\pi}
  \frac{d \phi_S}{2\pi}~
  e^{-i \, (B \phi_B + Q \phi_Q + S \phi_S)} \nonumber \\
  & \quad \times \exp\left[\sum_{j} z_j^1 \, e^{i \, (B_j \phi_B + Q_j \phi_Q + S_j \phi_S)}\right].
}
Here the sum denoted by index $j$ is over all hadrons and light nuclei included in the list. 
$B_j$, $Q_j$, and $S_j$ are, respectively, the baryon number, electric charge, and strangeness, carried by the particle species $j$, and $z_j^0$ is the one-particle partition function
\eq{\label{eq:zj}
z_j^1 = V_c \, \int dm \, \rho_j(m) \, d_j \frac{m^2 T}{2\pi^2} \, K_2(m/T),
}
where $d_j$ is the degeneracy factor for particle species $j$. 
The integration over the mass distribution $\rho_j(m)$ in Eq.~\eqref{eq:zj} takes into account the finite widths of the resonances.
In the present work we adopt the energy-dependent Breit-Wigner scheme, which was recently advocated for thermal model description at LHC energies~\cite{Vovchenko:2018fmh}.
The mean multiplicities of various particle species are calculated by introducing the fictitious fugacities into the partition function~\eqref{eq:Z} and calculating the corresponding derivatives with respect to these fugacities~(see details in Refs.~\cite{Becattini:1995if,Becattini:1997rv}).
The result is
\eq{
\langle N_j^{\rm prim} \rangle^{\rm ce} = \frac{Z(B-B_j,Q-Q_j,S-S_j)}{Z(B,Q,S)} \, \langle N_j^{\rm prim} \rangle^{\rm gce}~.
}
Here $\langle N_j^{\rm prim} \rangle^{\rm gce}$ are the mean multiplicities as calculated in the grand canonical ensemble at the same temperature $T$ and volume $V_c$, while the first factor is the chemical factor which appears due to the requirement of exact conservation of the conserved charges.
The final particle yields, $\langle N_j^{\rm tot} \rangle^{\rm ce}$, are then calculated after including the feeddown from the decays of unstable resonances~(see details in Ref.~\cite{Vovchenko:2018fmh}).

The crucial step is the numerical calculation of the canonical partition functions in Eq.~\eqref{eq:Z}.
This calculation can be tricky, especially at large volumes, where large cancellations between positive and negative contributions appear in the numerical integration. 
Simplifying schemes have been considered to make the calculation simpler.
In particular, in the case where no particle with baryon number $|B_j| > 1$ exists in the list, the integration over $\phi_B$ in \eqref{eq:Z} can be performed analytically~\cite{Keranen:2001pr}, greatly simplifying the numerics.
This is the option implemented in the \texttt{THERMUS} package~\cite{Wheaton:2004qb}.
The present work includes the light nuclei, with $|B_j| > 1$, in the list. 
Therefore, the method of Ref.~\cite{Keranen:2001pr} cannot be used here.
Instead, a direct full numerical integration is performed over all three charges in \eqref{eq:Z}  in the present work  using high-quality Gauss-Legendre quadratures.
The accuracy of the numerical calculations was cross-checked against several cases, where the analytic solution is known, as well as by observing a consistent approach of towards the grand-canonical limit, as the correlation volume $V_c$ is increased.

The particle list employed in the present work conservatively includes only the established hadrons and resonances listed in the 2014 edition of the Particle Data Tables~\cite{Agashe:2014kda}, as well as stable light nuclei up to $^4 {\rm He}$. 
Charm and bottom flavored hadrons are not considered.
The list of particles and their properties coincide with the ones used in Ref.~\cite{Vovchenko:2018fmh}, more details can be found there.

All calculations in this work are performed within the the \texttt{Thermal-FIST} package~\cite{ThermalFIST}, where the CSM as described above is implemented.

Matter produced at midrapidity at the LHC is practically net baryon free~\cite{Abelev:2013vea}.
Hence, the canonical ensemble HRG model is applied here for exactly vanishing values of the conserved charges, $B = Q = S = 0$.
Fits of the chemical equilibrium ideal HRG model employed here to hadron yields measured in central Pb-Pb collisions at the LHC consistently yield the chemical freeze-out temperature of about 155~MeV~\cite{Stachel:2013zma,Petran:2013lja,Floris:2014pta,Vovchenko:2018fmh},
hence we employ this value throughout the present ideal HRG analysis, unless stated otherwise.

\begin{figure*}[!ht]
  \centering
  \includegraphics[width=.47\textwidth]{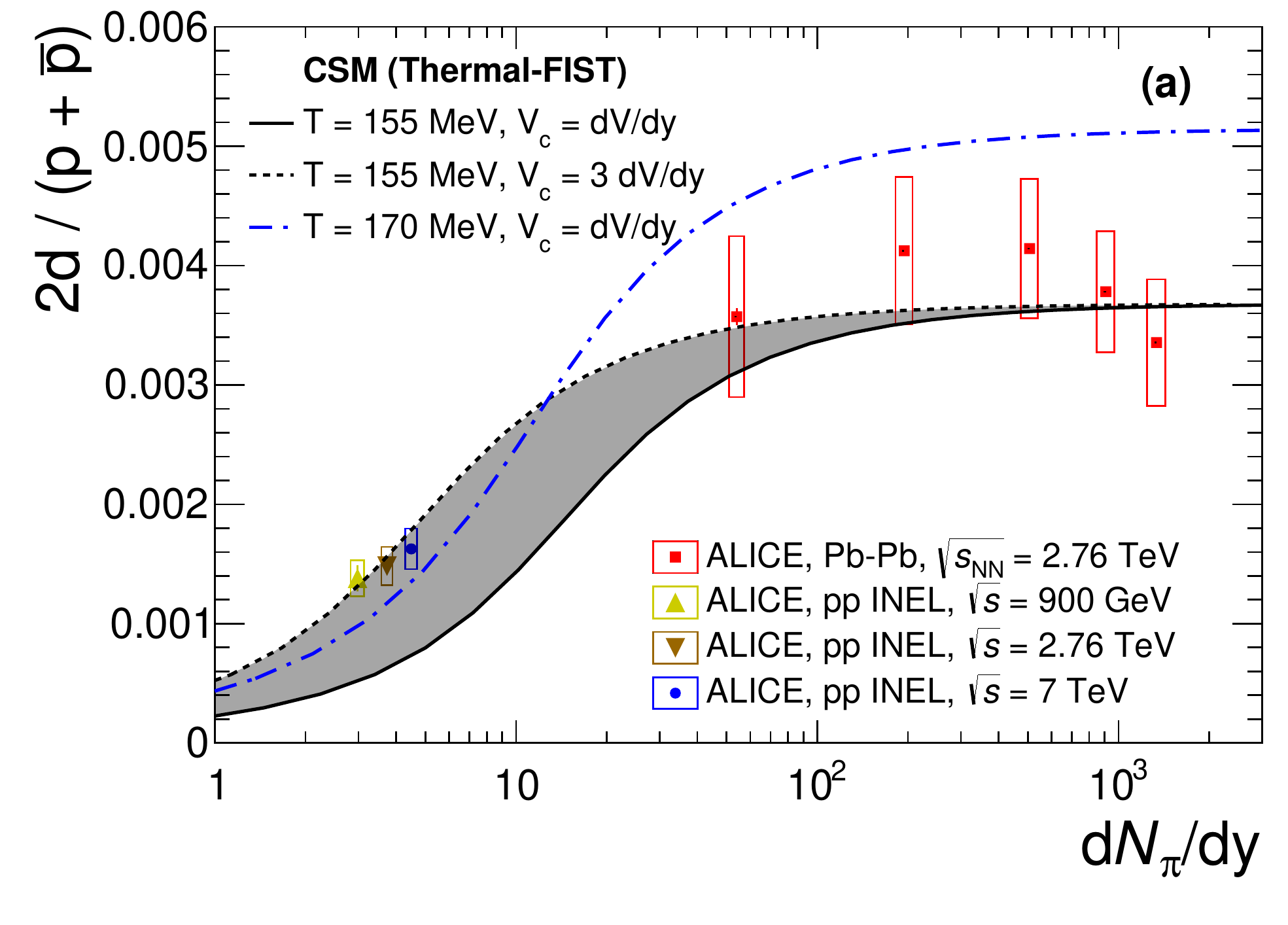}
  \includegraphics[width=.47\textwidth]{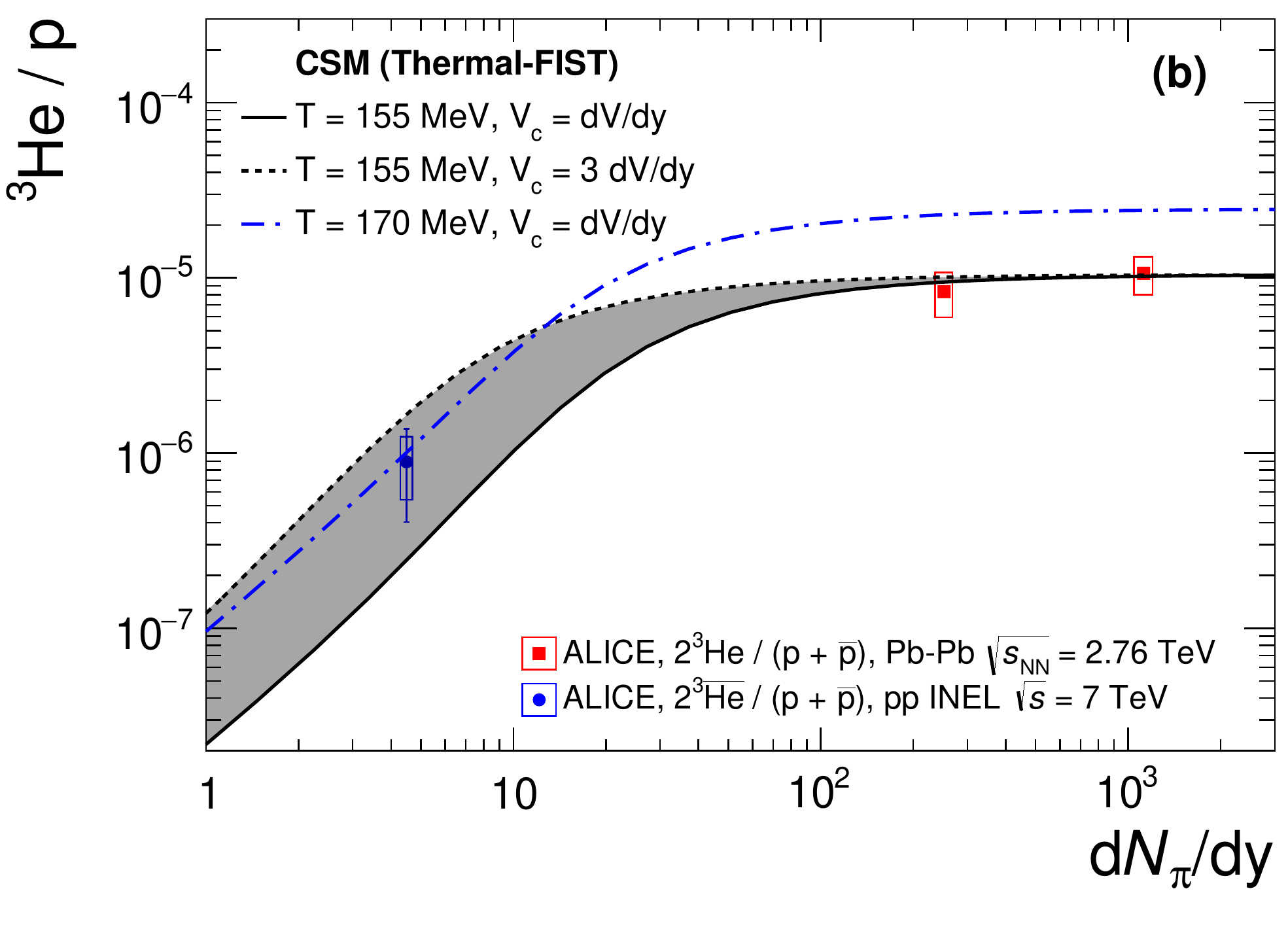}
  \includegraphics[width=.47\textwidth]{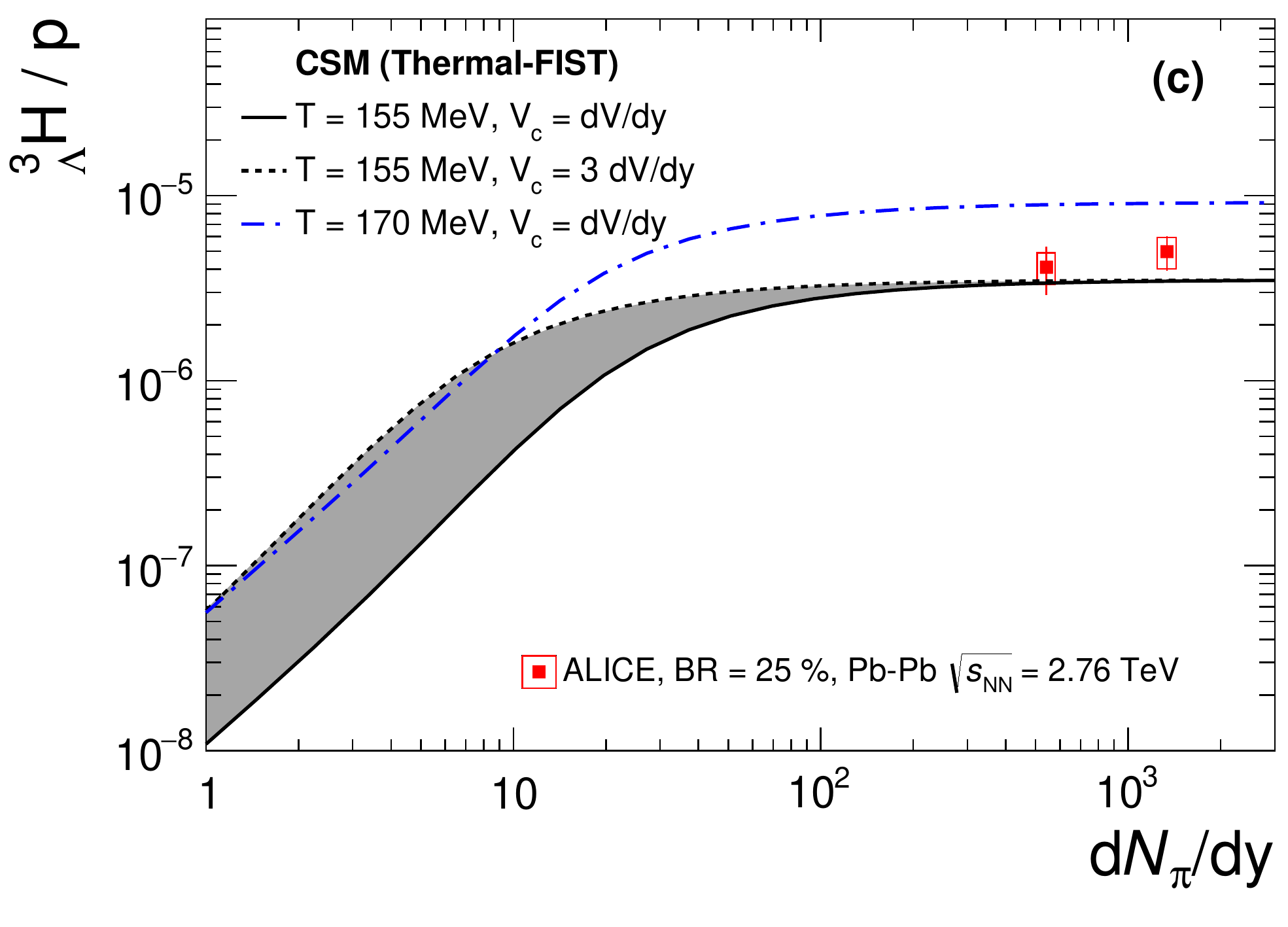}
  \includegraphics[width=.47\textwidth]{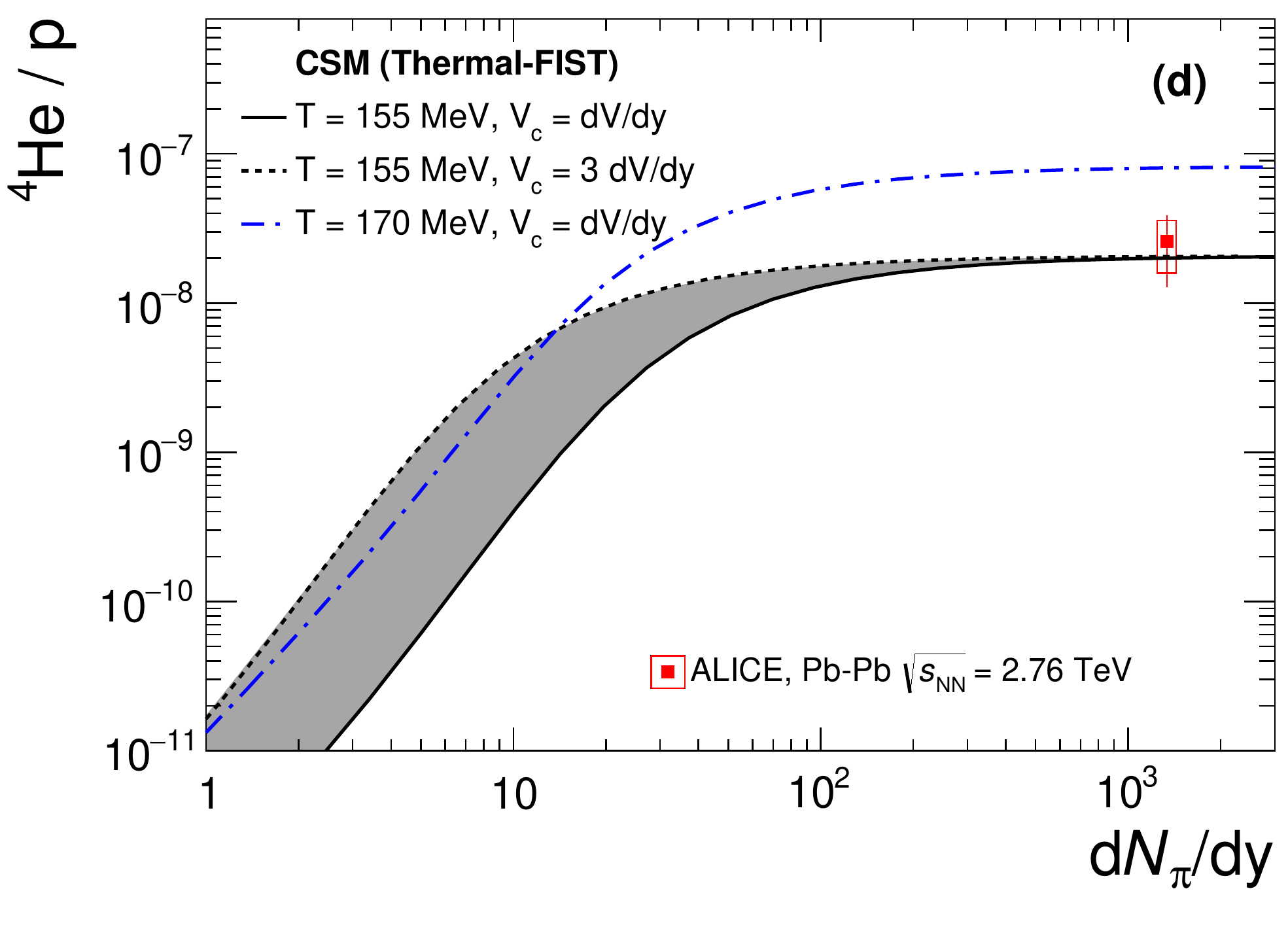}
  \caption{Charged pion multiplicity dependence of (a) $\text{d}/\text{p}$, (b) $^3\text{He} / \text{p}$, (c) $^3_{\Lambda} \text{H} / \text{p}$, and (d) $^4\text{He} / \text{p}$ ratios calculated in the canonical ensemble HRG model at $T = 155$~MeV for $V_c = dV / dy$~(solid black lines) and $V_c = 3 \, dV / dy$~(dashed black lines), and at $T = 170$~MeV for $V_c = dV / dy$~(dash-dotted blue lines).
  Experimental data of the ALICE collaboration~\cite{Adam:2015vda,Adam:2015yta,Acharya:2017bso,Acharya:2017fvb} are shown where available.
  }
  \label{fig:MultDep}
\end{figure*}

The multiplicity dependence is studied by varying the value of the correlation volume $V_c$.
To compare with the experiments,
we consider the dependence of the observables on the rapidity density $dN_{\pi} / dy$ of the charged pion multiplicity, which is measured in the experiments and can be calculated in the CSM.
To relate the mean multiplicities $\langle N^{\rm ce}_j \rangle^{\rm tot}$, as calculated in the CSM, to the rapidity densities $dN_j / dy$, the connection between the correlation volume $V_c$ and the volume $dV / dy$ corresponding to one unit of rapidity needs to be established.
As the midrapidity slice is an open system, where net values of conserved charges fluctuate from one event to another, there is no reason to enforce $V_c = dV / dy$.
It is also obvious that $V_c$ cannot exceed the total system volume.
Arguments based on the causal connection of fireballs populating the rapidity axis suggest that $V_c$ is smaller than the total volume and it may correspond to few units of rapidity~\cite{Castorina:2013mba}.
We thus vary $V_c$ in this work between $V_c = dV / dy$ and $V_c = 3 \, dV/dy$.
Consequently, one has $dN_j / dy = \langle N^{\rm tot}_j \rangle^{\rm ce}$ for $V_c = dV / dy$, and
$dN_j / dy = \langle N^{\rm tot}_j \rangle^{\rm ce} / 3$ for $V_c = 3 \, dV / dy$.
This variation has a considerable effect on the quantitative aspects of the results obtained.

Figure~\ref{fig:MultDep} depicts the charged multiplicity dependence of the following yield ratios calculated in the CSM at $T = 155$~MeV: (a) $\text{d}/\text{p}$, (b) $^3\text{He} / \text{p}$, (c) $^3_{\Lambda} \text{H} / \text{p}$, and (d) $^4\text{He} / \text{p}$.
The solid lines correspond to $V_c = dV / dy$, the dashed lines to $V_c = 3 \, dV / dy$. The shaded gray area corresponds to $dV / dy < V_c < 3 \, dV / dy$.
The additional CSM calculations at $T = 170$~MeV and $V_c = dV / dy$ are depicted by the dash-dotted blue lines.
The experimental data of the ALICE collaboration is shown in Fig.~\ref{fig:MultDep} by differently styled symbols.
These data include the $2 \text{d} / (\text{p} + \bar{\text{p}})$, $2 ^3\text{He} / (\text{p} + \bar{\text{p}})$, $^3_{\Lambda}\text{H} / p$, and $^4\text{He} / \text{p}$ ratios measured in Pb-Pb collisions at $\sqrt{s_{_{NN}}} = 2.76$~TeV~\cite{Adam:2015vda,Adam:2015yta,Acharya:2017bso}, the $2 \text{d} / (\text{p} + \bar{\text{p}})$ ratio measured in inelastic pp collisions at $\sqrt{s} = 0.9$, 2.76, and 7~TeV~\cite{Acharya:2017fvb}, and the $2 ^3\overline{\text{He}} / (\text{p} + \bar{\text{p}})$ ratio measured in inelastic pp collisions at $\sqrt{s} = 7$~TeV~\cite{Acharya:2017fvb}.
The $^3_{\Lambda}\text{H}$ yield was reconstructed from the ALICE data by assuming a 25\% branching ratio of the $^3_\Lambda\text{H} \to ^3 \text{He} + \pi$ decay~\cite{Adam:2015yta}.

All ratios show a monotonic increase with $dN_{\rm ch} / d \eta$, with a saturation at the corresponding grand-canonical values in the high-multiplicity limit.
These limiting grand-canonical values are in fair agreement with the experimental data for Pb-Pb collisions, as reported before~\cite{Adam:2015vda,Adam:2015yta,Acharya:2017bso}.
The available experimental data at smaller multiplicities presently includes only the $\text{d}/\text{p}$ and $^3 \text{He} / \text{p}$ ratios measured in inelastic pp interactions. 
These ratios are much  smaller than the grand-canonical limiting values.
The canonical suppression mechanism predicts a strong suppression at low multiplicities, consistent with the data.

The CSM underpredicts significantly for $T = 155$~MeV the pp data for the $\text{d}/\text{p}$ and $^3 \text{He} / \text{p}$ ratios for the case $V_c = dV / d y$.
A much better agreement is observed for $V_c \simeq 3 \, dV / dy$, indicating that the correlation volume extends across a few units of rapidity if the $T = 155$~MeV freeze-out temperature value is assumed constant across all systems.

Another possibility is the multiplicity dependent freeze-out temperature.
Recent CE fits to hadron yields measured in pp collisions at $\sqrt{s} = 7$~TeV do suggest freeze-out temperature values $T \gtrsim 170$~MeV for small systems~\cite{Sharma:2018uma}, higher than the typical $T \simeq 155$~MeV values for large systems created in Pb-Pb collisions.
The CSM results for the multiplicity dependence of the light nuclei-to-proton ratios calculated for $T = 170$~MeV and $V_c = dV / dy$ are shown in Fig.~\ref{fig:MultDep} by the dash-dotted blue lines.
The CSM at $T = 170$~MeV significantly overpredicts the high-multiplicity Pb-Pb values for all four ratios considered, but gives a fair description of the $\text{d}/\text{p}$ and $^3 \text{He} / \text{p}$ ratios measured in pp collisions.
This indicates a possibility that the chemical freeze-out temperature increases when going from high-multiplicity to low multiplicity events.

Energy-dependent Breit-Wigner scheme was employed to treat the finite resonance widths in the present work.
Finite widths have a negligible influence on the yields of light nuclei, since their final yields are virtually unaffected by feeddown at the LHC, 
but they do affect the proton yield due to a modified feeddown from broad $\Delta$ and $N^*$ resonances~\cite{Vovchenko:2018fmh}.
If the finite widths are neglected, then the CSM results in Fig.~\ref{fig:MultDep} would be similar, but pushed down by about 15\% due to an enhanced proton yield in the zero-width approximation relative to the energy dependent Breit-Wigner scheme.

Both the light nuclei and all hadrons were considered to be point-like in this study.
The validity of this assumption can be questioned.
The finite sizes of hadrons and light nuclei can be modeled through the excluded-volume correction~\cite{Rischke:1991ke}.
Recent studies indicate that a thermal description of particle yields as measured in Pb-Pb collisions at the LHC are sensitive to the modeling of the excluded-volume corrections~\cite{Vovchenko:2015cbk,Alba:2016hwx,Vovchenko:2016mwg}, thus leading to higher values of the chemical freeze-out temperatures.
The excluded volume effects on the multiplicity dependence of light nuclei yields can be studied in the canonical ensemble formulation of the excluded volume HRG model.
To our knowledge, such an analytic formulation is presently missing, although a recently developed Monte Carlo approach can be useful also in this respect~\cite{Vovchenko:2018cnf}.

To summarize,
the dependence of the light nuclei production at midrapidity on the charged pion multiplicity at LHC energies is considered here for the first time in the framework of the statistical model with exact conservation of baryon number, electric charge, and strangeness.
The ratios of the light nuclei-to-proton yields show a monotonic increase with the charged pion multiplicity, which saturates at the corresponding grand-canonical values in the high-multiplicity limit.
This result is in qualitative agreement with the experimental data measured by the ALICE collaboration in pp and Pb-Pb collisions for different centralities.
The experimental data for the $\text{d}/\text{p}$ and $^3 \text{He} / \text{p}$ ratios at low multiplicities suggest a canonical correlation volume which corresponds to exact conservation of charges across more than one unit of rapidity and/or a chemical freeze-out temperature which decreases with the charged pion multiplicity.


\begin{acknowledgments}

V.V. appreciates motivating discussions with the participants of the "Light up 2018 -- An ALICE and theory workshop", June 14-16, CERN, Switzerland.
B.D. acknowledges the support from BMBF through the FSP202 (F\"orderkennzeichen 05P15RFCA1).
H.St. acknowledges the support through the Judah M. Eisenberg Laureatus Chair by Goethe University  and the Walter Greiner Gesellschaft, Frankfurt.

\end{acknowledgments}

\bibliography{bibliography}


\end{document}